\documentclass[aip, apl, 10pt, twocolumn, reprint, superscriptaddress, amsmath, amssymb, showpacs, showkeys]{revtex4-1}
\usepackage{graphicx}
\usepackage{dcolumn}
\usepackage{bm}
\usepackage{upgreek}
\usepackage{times}
\usepackage{mathrsfs}
\usepackage{multirow}
\usepackage{fancyhdr} 
\usepackage{hyperref} 
\usepackage{color}

\begin{document}

\title{Valence band splitting in bulk dilute bismides}

\author{Lars C. Bannow}\email{lars.bannow@physik.uni-marburg.de}%
\affiliation{%
 Department of Physics and Material Science Center, Philipps-Universit\"at, 35032 Marburg, Germany
}%

\author{Stefan C. Badescu}
\affiliation{%
   Air Force Research Laboratory, Wright-Patterson AFB, Ohio 45433, USA
}%

\author{J{\"o}rg Hader}%
\affiliation{%
  Nonlinear Control Strategies Inc., 7040 N Montecatina Dr., Tucson, Arizona 85704, USA
}%

\author{Jerome V. Moloney}%
\affiliation{%
  Nonlinear Control Strategies Inc., 7040 N Montecatina Dr., Tucson, Arizona 85704, USA
}%

\author{Stephan W. Koch}%
\affiliation{%
 Department of Physics and Material Science Center, Philipps-Universit\"at, 35032 Marburg, Germany
}

\date{\today}

\begin{abstract}
The electronic structure of bulk GaAs$_{1-x}$Bi$_x$ systems for different atomic configurations and Bi concentrations is calculated using density functional theory. The results show a Bi-induced splitting between the light-hole and heavy-hole bands at the $\Gamma$-point. We find a good agreement between our calculated splittings and experimental data. The magnitude of the splitting strongly depends on the local arrangement of the Bi atoms but not on the uni-directional lattice constant of the supercell. The additional influence of external strain due to epitaxial growth on GaAs substrates is studied by fixing the in-plane lattice constants. \\ \\
{\small The following article has been submitted to Applied Physics Letters. If it is published, it will be found online at http://apl.aip.org}
\end{abstract}


\maketitle

%
%

The incorporation of relatively small concentrations of bismuth into semiconductor compounds is of growing interest because substitution of Bi not only allows for the engineering of the band gap $E_\text{g}$\cite{Francoeur_APL82_2003} of III-V semiconductors but also of the spin-orbit splitting $\Delta_\text{so}$\cite{Fluegel_PRL97_2006}. It has been proposed that making $\Delta_\text{so}$ larger than $E_\text{g}$ should lead to a decrease of the nonradiative Auger losses which is important to improve the output characteristics of semiconductor lasers\cite{Sweeney_ICTON_2011,Marko_JSTQE_23_2017}.

In addition to the intrinsic splitting between the spin-orbit split-off band (so) and the top-most valence bands, the degeneracy of the heavy-hole (hh) and light-hole (lh) bands at the $\Gamma$-point can also be lifted by symmetry reductions. A well-known example is the valence band splitting (VBS) in III-V semiconductors caused by strain due to lattice-mismatched growth \cite{Bir_1974}. Experimentally,
VBS was observed for GaAs$_{1-x}$Bi$_x$ layers grown lattice-mismatched on GaAs substrates\cite{Francoeur_APL82_2003, Batool_JAP111_2012, Alberi_PRB92_2015}. To analyze the observations, \citet{Batool_JAP111_2012} extracted the shear deformation potential for a Bi content up to $x = 10.4\,\%$. As previously shown for dilute nitrides \cite{Zhang_PRB61_2000}, the authors find that the shear deformation potential depends on the content of the substitute. However, in the case of GaAs$_{1-x}$N$_x$ the VBS shows a bowing which is absent for GaAs$_{1-x}$Bi$_x$. In addition, a VBS of bulk GaAs$_{1-x}$Bi$_x$ has been observed in tight-binding calculations by \citet{Usman_PRB84_2011} even though the authors do not further comment on it. They find an increase of the splitting in the case of compressive strain\cite{Usman_PRB87_2013} which slightly underestimates the splitting reported by \citet{Batool_JAP111_2012}. 

To systematically study Bi induced VBS, we use density functional theory (DFT) for the calculation of the band structures of dilute bismide semiconductors. Remarkably, our results show that a large VBS in the range of tens to more than hundred meV exists for bulk systems without any external strain. The magnitude of the splitting strongly depends on the local atomic arrangement of the Bi atoms in the GaAs lattice. Similar cluster-induced VBS has previously also been reported for GaAs$_{1-x}$N$_x$\cite{Kent_PRB64_2001}. For a given atomic configuration, the VBS can be further increased when a lattice-mismatched growth is modeled by allowing the lattice constant in growth direction to relax while fixing the in-plane lattice constant to the GaAs one.

%
%

The density functional theory calculations were performed with the \textit{Vienna Ab-initio Simulation Package}~\cite{Kresse_PRB_47_1993,Kresse_PRB_49_1994,Kresse_CMS_6_1996,Kresse_PRB_54_1996} (VASP) and the Projector Augmented-Wave (PAW) pseudopotential method~\cite{Bloechl_PRB_50_1994,Kresse_PRB_59_1999} was applied. For optimization, we used an energy-cutoff of $510\,$eV, the GGA-PBEsol functional~\cite{Perdew_PRL_100_2008}, Hellmann-Feynman forces on the atoms were minimized below $20\,$meV/{\AA}, and the energies were converged to an accuracy of $10^{-6}\,$eV, respectively. When calculating the band energies, spin-orbit coupling was enabled, an energy-cutoff of $410\,$eV was applied, the energies were converged to an accuracy of $10^{-4}\,$eV and the Tran-Blaha modified Becke Johnson potential (TBmBJ)~\cite{Tran_PRL_102_2009} was used. This potential yields close-to-experiment band gaps for III-V semiconductor materials while being computationally efficient.
The $k$-space was sampled with an $8\times 8 \times 8$ Monkhorst-Pack mesh~\cite{Monkhorst_PRB_13_1976} in the case of primitive cells (2 atoms) and with a $2 \times 2 \times 2$ Monkhorst-Pack mesh in the case of 128 atom supercells, respectively.

As a reference, we note that we obtain a lattice constant of $a_0 = 5.653\,${\AA} for pure GaAs, a band gap of $E_g = 1.50\,$eV and a spin-orbit splitoff of $\Delta_{\textnormal{SO}} = 0.31\,$eV, respectively. A VBS between the lh and hh band is absent ($0\,$meV) in this case.

In several of our calculations for different Bi concentrations, we used the \textit{Alloy-theoretic automated toolkit} (ATAT)\cite{Walle_Calphad26_2002,Walle_Calphad33_2009} to generate special quasirandom structures (SQS)\cite{Zunger_PRL65_1990, Walle_Calphad42_2013} for 128 atom supercells. SQS are the best approximation of the true disordered state when the supercell size is limited. In this work, the pair length includes third nearest neighbors, the triplet length includes second nearest neighbors and the quadruplet length includes nearest neighbors, respectively.

In addition to calculations where the volume of the GaAs$_{1-x}$Bi$_x$ supercell was fixed to the volume of a GaAs supercell with an equal amount of atoms (denoted by \textit{strain}), \textit{relaxed} GaAs$_{1-x}$Bi$_x$ supercells were created (denoted by \textit{relax}).
The relaxed supercells were obtained by minimizing the total energy $E_\text{tot}$ as a function of the global lattice constant $a_0$ ($a_0^{001}$ in [001] direction for \textit{on substrate} modeling). We used a quadratic function to fit $E_\text{tot}(a_0)$ and extracted the lattice constant that minimizes $E_\text{tot}$.

\begin{figure}[t]
\includegraphics[width=8.5cm]{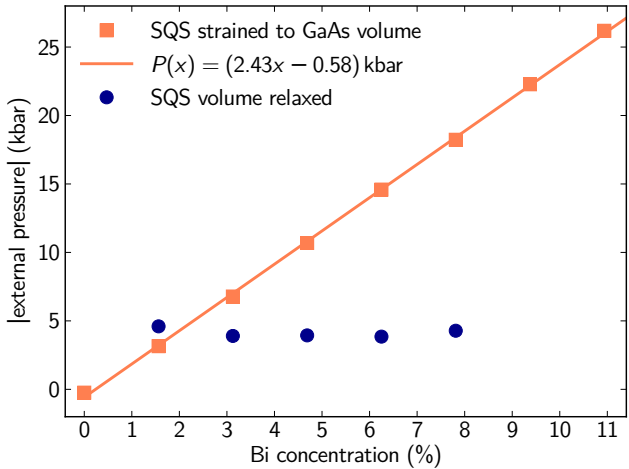}
\caption{The absolute value of the external pressure on the supercell as a function of Bi concent in the supercell. The orange data shows the pressure of the strained supercells whereas the blue data shows the pressure of the relaxed supercells.}
\label{fig:figD}
\end{figure}

For illustration, we plot in Fig.~\ref{fig:figD} the external pressure on the supercell as a function of the Bi content. Only the pressure for the SQS is shown but other 128-atom supercells exhibit a similar behavior. It can be seen that the pressure of the strained supercells increases linearly with increasing Bi content (increasing amount of Bi atoms in the supercell) by about $2.43\,$kbar\,/\,$\%$Bi. The origin of this is the larger size of Bi atoms when compared to the As atoms which they substitute. For the relaxed supercell the pressure remains almost constant slightly above $-5\,$kbar. This is most likely due to (a) forcing the lattice constant in $x$-, $y$- and $z$-direction to be equal and (b) enforcing a lattice symmetry.

%
%

\begin{figure}[t]
\includegraphics[width=5cm]{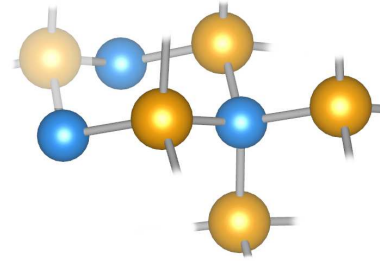}
\caption{Arrangement of Bi atoms (orange) and Ga atoms (blue) in a five Bi atom cluster. In case of a four Bi atom cluster the leftmost greyed out Bi atom is replaced by an As atom. Additionally, for a three (two) Bi atom cluster, one (two) more Bi atoms are removed. Only bonds between Bi and Ga atoms are shown.}
\label{fig:figA}
\end{figure}

To complement the analysis of the SQS configurations, the band gap $E_\text{g}$ and the VBS $E_\text{VBS}$ were also calculated for clustered arrangements of Bi atoms in a $4 \times 4 \times 4$ (128 atom) GaAs$_{1-x}$Bi$_x$ supercell. In these clustered arrangements all Bi atoms are located in the vicinity of the same Ga atom (see Fig.~\ref{fig:figA}).
 
The calculated values of the band gap and VBS for strained and relaxed supercells are summarized in Tab.~\ref{Table:A} together with the lattice constant for the relaxed supercells. As we already showed in Ref.~\cite{Bannow_PRB93_2016}, the band gap narrowing depends strongly on the arrangement of the Bi atoms in the supercell. We demonstrated that the reason for this is a different overlap among the Bi \textit{p} orbitals for each arrangement. Relaxing the volume of the supercell enhances the narrowing of the band gap with increasing Bi concentration by roughly a factor of $1.5$ (cf. Tab.~\ref{Table:A}). This is expected since an increase of the lattice constant alters the edges of the first Brillouin zone. The computed band gap narrowing of the SQS ($72\,$meV/$\,\%$Bi) is in good agreement with experimental results ($60-70\,\text{meV}$)\cite{Breddermann_JL_154_2014}.

\begin{table*}
    \caption{Band gap $E_\text{g}$ and VBS $E_\text{VBS}$ for different arrangements of one, two, three, four and five Bi atoms in 128 atom supercells. Results are shown for two cases: (a) the lattice constant of the supercell was fixed to the one of GaAs in all three directions (strain) and (b) the volume of the supercell was relaxed to account for the incorporation of the larger Bi atom in the host lattice (relax). Since the results of different SQS with the same Bi content can differ significantly, we show the range for the strained supercell. For the relaxed supercell, no range is included because we only calculated the data for one specific SQS at each concentration. In the case of the clustered arrangement the data for one Bi atom has been put in brackets since this arrangement is not a cluster and only serves as a reference.}\label{Table:A}
    \begin{ruledtabular}
        \begin{tabular}{l c c c c c c c c}
            &Arrangement & 1 atom & 2 atoms & 3 atoms & 4 atoms & 5 atoms & $\Delta \frac{E}{x}~\left(\frac{\mathsf{meV}}{\mathsf{\%Bi}}\right)$  \\
            \hline
            &bulk clustered& \{(0,0,0)\}&\{(0,0,0), & \{(0,0,0), &  \{(0,0,0),& \{(0,0,0),&    \\
            &         &           & (1,0,0)\} & (1,0,0),   &  (1,0,0),  & (1,0,0), &      \\
            &         &           &           & (0,1,0)\}  &  (0,1,0),  & (0,1,0), &      \\
            &         &           &           &            &  (0,0,1)\} & (0,0,1), &      \\
            &         &           &           &            &            & (1,1,0)\}&      \\[.4ex]
$E_\text{g}^\text{strain}$ (eV)   &           & (1.37)     &  1.18      & 1.02     & 0.96 & 0.86 & $-84\pm7$   \\
$E_\text{VBS}^\text{strain}$ (eV)&           & (0)        &  0.14      & 0.18     & 0    & 0.10 & $-$        \\ [.4ex]
$E_\text{g}^\text{relax}$ (eV)    &           &  (1.32)    &  1.08      & 0.88     & 0.77 & 0.63  & $-120\pm10$ \\
$E_\text{VBS}^\text{relax}$ (eV) &           &  (0)       &  0.14      & 0.17     & 0    & 0.11  & $-$        \\
$a_0^\text{relax}$ (\AA)          &           & (5.672)      & 5.680      & 5.693    & 5.706     & 5.718 & \\
             \hline
            &bulk SQS     &         &          &         &         &        &                   \\[.4ex]
$E_\text{g}^\text{strain}$ (eV)&        & 1.43...1.44    & 1.37...1.38     & 1.24...1.31    & 1.19...1.25    & 1.15...1.22   &  $-42\pm 3$&  \\
$E_\text{VBS}^\text{strain}$ (eV)&& 0.01    & 0.01...0.04     & 0.04...0.06    & 0.05...0.13    & 0.04...0.13   &         $-$  \\ [.4ex] 
$E_\text{g}^\text{relax}$ (eV)&         &  1.36 &  1.26      & 1.11       &  1.06     & 0.92 &  $-72\pm4$  & \\
$E_\text{VBS}^\text{relax}$ (eV)&      &  0.01    &  0.03         & 0.06          &  0.09        & 0.13    &  $-$  \\ 
$a_0^\text{relax}$ (\AA)         &            & 5.674   & 5.682      & 5.692      & 5.702     & 5.714 & \\
        \end{tabular}
    \end{ruledtabular}
\end{table*}
In our calculations with finite Bi concentrations, we observe a certain amount of VBS for the clustered and SQS arrangements. Interestingly, we find no significant changes when we relax the volume of the supercell as compared to the strained case. For the SQS arrangements, the general trend is that the splitting increases with Bi concentration. However, the absolute value of the VBS strongly depends on the specific arrangement of the Bi atoms in the SQS. In other words, the VBS for different SQS with the same Bi content can vary significantly; we have found differences of up to $90\,$meV for one and the same Bi content. 

\begin{figure}
\includegraphics[width=8.5cm]{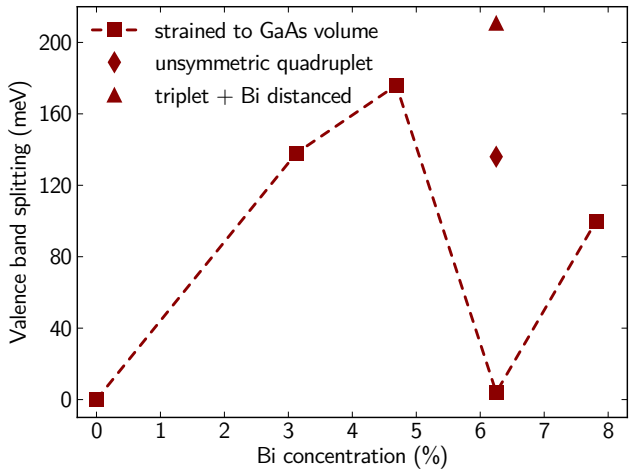}
\caption{Valence band splitting for clustered arrangements. The Bi atoms are arranged such that they are neighbors to the same Ga atom (squares), three as neighbors to the same Ga atom and one far away from the cluster in the supercell (diamond) and such that three are neighbors to the same Ga atom and a fourth Bi is arranged such that it shares a common Ga neighbor with only one of the other 3 Bi atoms (diamond).}
\label{fig:figB}
\end{figure}

Surprising results are also found for the clustered arrangements: For four Bi atoms that are all neighbors to the same Ga atom (see Fig.~\ref{fig:figA}), there is no VBS. In a very similar arrangement, however, where we moved one Bi atom to a slightly different position so that it is a next-nearest neighbor of the center Ga atom (in Fig.~\ref{fig:figA} this corresponds to moving the left-most Bi atom that is not grayed-out to the position of the grayed-out Bi atom) the splitting is roughly $135\,$meV. Moving that atom far away from the cluster further increases the splitting to above $200\,$meV. Lastly, when a fifth Bi atom is added to the symmetric quadruplet arrangement (see Fig.~\ref{fig:figA}), the resulting splitting is smaller than for the other clustered cases. A summary of the computed VBS for the different clustered arrangements is presented in Fig.~\ref{fig:figB}.

The observed VBS behavior suggests that the splitting is caused by local symmetry breaking within the lattice. Very minimal splitting occurs for a cluster of four Bi atoms since this is a locally symmetric arrangement. In contrast, all the other clustered arrangements and the SQS represent locally unsymmetric arrangements causing the observed splitting. 


\begin{figure}
\includegraphics[width=8.55cm]{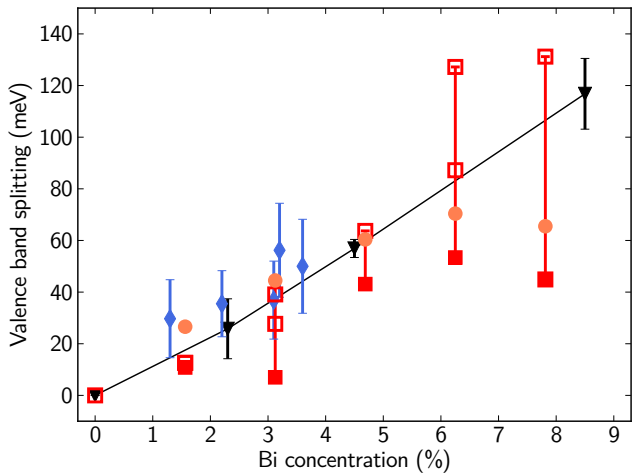}
\caption{Comparison of our SQS VBS results to experimental data. In the experiments the GaAs$_{1-x}$Bi$_x$ layer was grown on a GaAs substrate. The data marked with black triangles is taken from \citet{Batool_JAP111_2012}. The line serves as a guide to the eye. The data marked with blue diamonds is taken from ~\citet{Francoeur_APL82_2003}. The red squares show the splittings obtained for two or three SQS configurations at the same Bi concentration and the error bars show the corresponding range. The filled orange circles show the splitting of $xy$-strained SQS supercells and can be compared to the filled red squares that show the VBS of the corresponding volume strained SQS supercell.}
\label{fig:figC}
\end{figure}

It is expected that the SQS represent atomic configurations that are closest to experimental systems since they contain clustering as well as regions with sparse Bi content. Therefore, we compare our SQS splitting results from Tab.~\ref{Table:A} to splittings measured by photoreflectance spectroscopy \cite{Batool_JAP111_2012,Francoeur_APL82_2003}. In the experiments, the GaAs$_{1-x}$Bi$_x$ layer was grown on a GaAs substrate leading to strain in the $xy$-plane. To account for this feature, we constructed some SQS supercells where the lattice constant in $z$-direction was relaxed with the same method that was used for the volume relaxation while the lattice constants in $x$- and $y$-direction were fixed to the GaAs lattice constant of $a_0 = 5.653\,$\AA. In Fig.~\ref{fig:figC}, these are labeled \textit{SQS-128 xy-strained} and represented by filled orange circles. The results for the corresponding uniformly strained supercells are shown as filled red squares.

We notice, that the $xy$-strained SQS show a larger VBS than that of the matching uniformly strained supercell. This is expected since the relaxation in $z$-direction lifts the symmetry. 
Generally, up to a Bi concentration of about $6\,\%$ the VBS increases. However, as already mentioned earlier, for the same Bi concentration, the VBS of the SQS can strongly divert. For the same degree of overall randomness, this is caused by different local arrangements of the Bi atoms. Generally,  larger VBS occurs as soon as locally clustered configurations are contained in the supercell.

Next, we compare our DFT result with experimental data. In the Bi concentration range up to $6\,$\% all our calculations resemble the experimental data within a range that is comparable to the experimental error. For larger Bi contents the range of observed VBS in our calculations is larger than the experimental error but still covers the range of the experimental results.

To check the general aspects of our observations, we repeated some of the calculations also for InAs$_{1-x}$Bi$_x$ and obtained similar results. Since VBS has also been reported in theoretical studies for bulk GaAs$_{1-x}$N$_x$\cite{Kent_PRB64_2001}, we expect that it also occurs in other III-V compounds. Hence, we want to note that VBS caused by the lifting of the local lattice symmetry needs to be accounted for when calculating the shear derformation potential from experimentally measured splittings.

%
%

In conclusion, our DFT calculations for bulk GaAs$_{1-x}$Bi$_x$ supercells yield a VBS without any external symmetry reduction. The splitting increases with the Bi concentration and strongly depends on the local arrangement of the Bi atoms in the supercell. For the same Bi concentration and different arrangements we observed splittings that differ by more than $200\,$meV in the most extreme case (see Fig.~\ref{fig:figB}). We show that the VBS is caused by a distortion of the lattice symmetry due to the incorporation of the Bi atoms. The splitting can be further increased when the supercell is relaxed in $z$-direction while keeping the in-plane lattice constants fixed to the GaAs lattice constant. The reason for this is an additional lifting of the cubic lattice symmetry. Since in experiments the GaAs$_{1-x}$Bi$_x$ layer is usually grown lattice-mismatched on a GaAs substrate the latter case compares best to experiment. We find a good agreement between our results for SQS and experimental results for the splitting.

%
%
\begin{acknowledgments}
The Marburg work was funded by the DFG via the GRK 1782 "Functionalization of Semiconductors"; computing time from HRZ Marburg, CSC Frankfurt and HRZ Darmstadt is acknowledged. The work at Nonlinear Control Strategies Inc. was supported by the Air Force Office of Scientific Research under the STR Phase II grant FA9550-16-C-0021. The work of SCB was supported by the AFOSR and by the HPC of the DoD.

Fig.~\ref{fig:figA} was created with VESTA\cite{Momma_JApplCryst_44_2011}.
\end{acknowledgments}

%
%
\bibliography{lhhhlit}

\end{document}